\newcommand{\be}{\begin{equation}}
\newcommand{\ee}{\end{equation}}
\newcommand{\bea}{\begin{eqnarray}}
\newcommand{\eea}{\end{eqnarray}}
\newcommand{\bc}{\left\{\begin{aligned}}
\newcommand{\ec}{\end{aligned}\right.}
\newcommand{\rd}[1]{\,{\rm d}#1}
\newcommand{\ri}{{\rm i}}
\newcommand{\re}[1]{{\rm e}^{#1}}
\newcommand{\vv}[1]{\boldsymbol{\mathrm{#1}}}
\newcommand{\hvv}[1]{\boldsymbol{\hat{\mathrm{#1}}}}
\newcommand{\pint}{\:\mathcal{P}\!\!\int}
\newcommand{\eref}[1]{(\ref{#1})}
\newcommand{\Eref}[1]{Equation (\ref{#1})}
\newcommand{\fref}[1]{Fig.~\ref{#1}}

\newcommand{\sref}[1]{\S \ref{#1}}

\documentclass[osajnl,twocolumn,showpacs,superscriptaddress,10pt]{revtex4-1} 
\usepackage{amsmath,amssymb,graphicx}
\usepackage{color}
\newcommand{\mm}[1]{#1}

\begin{document}

\title{\mm{Singular eigenfunctions for the three-dimensional 
radiative transport equation}}

\author{Manabu Machida}\email{mmachida@umich.edu}
\affiliation{Department of Mathematics, 
University of Michigan, Ann Arbor, MI 48109, USA}

\begin{abstract}
Case's method obtains solutions to the radiative transport equation as 
superpositions of elementary solutions when the specific intensity depends 
on one spatial variable.  In this paper, we find elementary solutions when 
the specific intensity depends on three spatial variables in three-dimensional 
space.  
\mm{By using the reference frame whose $z$-axis lies in the direction 
of the wave vector, the angular part of each elementary solution becomes 
the singular eigenfunction for the one-dimensional radiative 
transport equation.}  Thus Case's method is generalized.
\end{abstract}


\maketitle 

\section{Introduction}

We consider light propagating in random media such as 
\mm{fog, cloud, and }
biological tissue.  
Then the specific intensity of light obeys the radiative transport equation.  
\mm{Although different numerical methods have been developed 
\cite{Case-Zweifel,Duderstadt-Martin,Marchuk-Lebedev}, analytical approach 
is preferable particularly for the sake of medical imaging and optical 
tomography \cite{Arridge99,Arridge09}.
}

Case's method is a method of obtaining solutions to the equation as 
superpositions of elementary solutions \cite{Case60}.  
Although the method gives insight into the theoretical structure of the 
specific intensity, it works \mm{only} when the specific intensity 
carries one spatial variable and is independent of two spatial variables 
in three-dimensional space.  
\mm{While the extension of Case's method to 
anisotropic scattering was soon done \cite{Mika61,McCormick-Kuscer66}, 
there has been no real success in extending the method to three dimensions 
despite considerable efforts 
\cite{Williams68,Kaper69,Gibbs69,Garrettson-Leonard70,
Case-Hazeltine70,Cannon73,Pomraning96}.  }
In particular, Kaper proposed \mm{elementary solutions of the form of a 
plane wave and developed a singular-eigenfunction theory by reducing the 
problem to a one-dimensional equation by changing angular variables to a 
new complex variable \cite{Kaper69}.  
However, this singular-eigenfunction is complicated (for example, 
the dispersion function $\Lambda$ is given by a three-dimensional integral) 
\cite{Case-Hazeltine70}.  Even for an infinite medium with isotropic 
scattering, calculation is quite complicated.  
Duderstadt and Martin wrote ``Although there have been many attempts to extend 
these methods (the integral transform and singular eigenfunction methods) to 
two- and three-dimensional problems, these extensions have usually encountered 
extreme mathematical complexity and have met with only marginal success.'' 
(\cite{Duderstadt-Martin}, p.~122).
}

In this paper, we extend Case's method to a general case where the 
specific intensity depends on three spatial variables in addition to 
two angular variables.  We evaluate the singular eigenfunction in each 
elementary solution with the reference frame whose $z$-axis is taken in 
the direction of the wave vector.  That is, the reference frame is rotated 
depending on the transverse buckling constants.  
\mm{This point is the key difference from Kaper's singular eigenfunctions.  
Indeed, the idea of rotated reference frames was first used 
by Markel\cite{Markel04}; 
the angular part of elementary solutions was expanded by 
rotated spherical harmonics.  
}

The \mm{remainder of the} paper is organized as follows.  In \sref{rte}, we 
introduce the radiative transport equation.  In \sref{modes}, we 
\mm{develop singular eigenfunctions and obtain elementary solutions}.  
In \sref{evals}, we consider eigenvalues.  We see the relation to 
the method of rotated reference frames in \sref{mrrf}.  
In \sref{gfunc3d}, we obtain the three-dimensional Green's function in 
an infinite medium.  Then the energy density is calculated as a numerical 
example in \sref{den}.  Finally, we give summary in \sref{summary}.  
Polar and azimuthal angles in rotated reference frames are presented in 
Appendix \ref{rotmuphi}.  The expansion coefficients in the 
method of rotated reference frames are calculated in Appendix \ref{efunc}.

\section{Radiative transport equation}
\label{rte}

Let $\mm{I}(\vv{r},\hvv{s})$ be the specific intensity at position
$\vv{r}\in\mathbb{R}^3$ in direction $\hvv{s}\in\mathbb{S}^2$.
We consider the time-independent radiative transport equation, 
which is given by
\mm{
\bea
\hvv{s}\cdot\nabla I(\vv{r},\hvv{s})
+(\mu_{\rm a}+\mu_{\rm s})I(\vv{r},\hvv{s})
&=&
\mu_{\rm s}\int_{\mathbb{S}^2}f(\hvv{s}\cdot\hvv{s}')\mm{I}(\vv{r},\hvv{s}')
\rd{\hvv{s}'}
\nonumber \\
&+&
S(\vv{r},\hvv{s}),
\label{rte:fullrte1}
\eea
where $\mu_{\rm a}$ and $\mu_{\rm s}$ are the absorption and scattering
coefficients, respectively, and $S(\vv{r},\hvv{s})$ is the source term.
We suppose $\mu_{\rm a}$ and $\mu_{\rm s}$ are positive constants, and 
}
the scattering phase function $f(\hvv{s}\cdot\hvv{s}')$
can be modeled by a polynomial of spherical harmonics of order $N$:
\be
f(\hvv{s}\cdot\hvv{s}')
=\sum_{l=0}^N\sum_{m=-l}^lf_lY_{lm}(\hvv{s})Y_{lm}^*(\hvv{s}').
\label{phasefunc}
\ee
We
\mm{choose $f_l$ so that $f$ is normalized as}
\be
\int_{\mathbb{S}^2}f(\hvv{s}\cdot\hvv{s}')\rd{\hvv{s}'}=1.
\ee
We have $N=0$, $f_0=1$ in the case of isotropic scattering, and 
have $N=1$, $f_0=1$, 
$f_1=\int_{\mathbb{S}^2}(\hvv{s}\cdot\hvv{s}')f(\hvv{s}\cdot\hvv{s}')
\rd{\hvv{s}'}$ in the case of linear scattering.  For the Henyey-Greenstein 
model \cite{Henyey-Greenstein41}, we have $N=\infty$, $f_l=\mm{f_1^l}$.  
\mm{By dividing both sides of \eref{rte:fullrte1} by 
$\mu_{\rm t}=\mu_{\rm a}+\mu_{\rm s}$, we obtain
\bea
\hvv{s}\cdot\nabla_{\tilde{\vv{r}}}I(\tilde{\vv{r}}/\mu_{\rm t},\hvv{s})
+I(\tilde{\vv{r}}/\mu_{\rm t},\hvv{s})
&=&
c\int_{\mathbb{S}^2}f(\hvv{s}\cdot\hvv{s}')I(\tilde{\vv{r}}/\mu_{\rm t},\hvv{s}')
\rd{\hvv{s}'}
\nonumber \\
&+&
\frac{1}{\mu_{\rm t}}S(\tilde{\vv{r}}/\mu_{\rm t},\hvv{s}),
\label{rte:fullrte1.5}
\eea
where $c=\mu_{\rm s}/\mu_{\rm t}$ is a constant, $0<c<1$, and 
$\tilde{\vv{r}}=\mu_{\rm t}\vv{r}$.  By writing
\be
\tilde{I}(\tilde{\vv{r}},\hvv{s})
=I(\tilde{\vv{r}}/\mu_{\rm t},\hvv{s}),
\ee
we obtain
\be
\hvv{s}\cdot\nabla_{\tilde{\vv{r}}} \tilde{I}(\tilde{\vv{r}},\hvv{s})
+\tilde{I}(\tilde{\vv{r}},\hvv{s})=
c\int_{\mathbb{S}^2}f(\hvv{s}\cdot\hvv{s}')\tilde{I}(\tilde{\vv{r}},\hvv{s}')
\rd{\hvv{s}'}+\frac{1}{\mu_{\rm t}}S(\tilde{\vv{r}}/\mu_{\rm t},\hvv{s}).
\label{rte:fullrte2}
\ee
Hereafter we will take the unit of length to be $1/\mu_{\rm t}$ and drop 
the tilde ``$\tilde{\phantom{I}}$''.
}

The specific intensity $I$ in
\eref{rte:fullrte2} is given as a superposition of elementary solutions, 
which are solutions to the following homogeneous equation.
\be
\hvv{s}\cdot\nabla \mm{I}(\vv{r},\hvv{s})+\mm{I}(\vv{r},\hvv{s})=
c\int_{\mathbb{S}^2}f(\hvv{s}\cdot\hvv{s}')\mm{I}(\vv{r},\hvv{s}')
\rd{\hvv{s}'}.
\label{rte:rte}
\ee

Let $\mu=\cos\theta$ be the cosine of the polar angle of 
$\hvv{s}$ and $\varphi$ be the azimuthal angle of $\hvv{s}$.  
\mm{Following \cite{McCormick-Kuscer66}, we} express $f(\hvv{s}\cdot\hvv{s}')$ 
in \eref{phasefunc} as
\bea
f(\hvv{s}\cdot\hvv{s}')
&=&
\sum_{l=0}^N\sum_{m=-l}^lf_l\frac{2l+1}{4\pi}\frac{(l-m)!}{(l+m)!}
\left(1-\mu^2\right)^{|m|/2}
\nonumber \\
&\times&
\left(1-{\mu'}^2\right)^{|m|/2}
p_l^m(\mu)p_l^m(\mu')\re{\ri m(\varphi-\varphi')}.
\eea
Here the polynomials $p_l^m(\mu)$ are related to associated Legendre 
polynomials $P_l^m(\mu)$ as 
\cite{McCormick-Kuscer66}
\be
P_l^m(\mu)=(-1)^m\left(1-\mu^2\right)^{|m|/2}p_l^m(\mu).
\ee
They satisfy the following recurrence relations and orthogonality relations.
\be
(l-m+1)p_{l+1}^m(\mu)=(2l+1)\mu p_l^m(\mu)-(l+m)p_{l-1}^m(\mu),
\label{prec}
\ee
\be
\int_{-1}^1p_l^m(\mu)p_{l'}^m(\mu)\rd{m(\mu)}=
\frac{2(l+m)!}{(2l+1)(l-m)!}\delta_{ll'},
\ee
where we introduced
\be
\rd{m(\mu)}=\left(1-\mu^2\right)^{|m|}\rd{\mu}.
\ee
\mm{Furthermore we have
\be
p_{|m|}^m(\mu)=
\bc
&\frac{(2m)!}{2^mm!}&\quad\mbox{for}\quad m\ge0,
\\
&\frac{(-1)^m}{2^{|m|}(|m|!)}&\quad\mbox{for}\quad m<0.
\ec
\ee
}

\section{\mm{Elementary solutions}}
\label{modes}

We seek solutions of the form of plane-wave decomposition 
\cite{Kaper69,Kim-Keller03,Kim04}.  We introduce $\nu\in\mathbb{R}$ 
and $\vv{q}\in\mathbb{R}^2$, and define vector $\vv{k}\in\mathbb{C}^3$ as
\be
\vv{k}=\frac{1}{\nu}\hvv{k},\quad
\hvv{k}=\left(\begin{array}{c}
-\ri\nu\vv{q} \\ Q(\nu \mm{q})
\end{array}\right),\quad
Q(\nu \mm{q})=\sqrt{1+\mm{(\nu q)^2}},
\ee
\mm{where $q=|\vv{q}|$.}  
We emphasize that $\vv{k}$ and $\hvv{k}$ are functions of $\nu$ and $\vv{q}$.  
We assume the specific intensity of the form
\be
\mm{I}_{\nu}^m(\vv{r},\hvv{s};\vv{q})
=\Phi^m_{\nu}(\hvv{s};\hvv{k})\re{-\vv{k}\cdot\vv{r}},
\label{modes:ansatz}
\ee
where
\be
\Phi^m_{\nu}(\hvv{s};\hvv{k})
=\phi^m(\nu,\mu(\hvv{k}))\left(1-\mu(\hvv{k})^2\right)^{|m|/2}
\re{\ri m\varphi(\hvv{k})}.
\label{modes:ansatz2}
\ee
Here $\mu(\hvv{k})$ and $\varphi(\hvv{k})$ are the cosine of the polar 
angle of $\hvv{s}$ and the azimuthal angle of $\hvv{s}$, respectively, 
in the rotated reference frame whose $z$-axis coincides with the direction of 
$\hvv{k}$ (see Appendix \ref{rotmuphi}).  
\mm{Note that in the laboratory frame 
($\hvv{k}=\hvv{z}$), \eref{modes:ansatz} reduces to the form used in 
\cite{McCormick-Kuscer66}.}  
We 
\mm{will determine elementary solutions 
$\mm{I}_{\nu}^m(\vv{r},\hvv{s};\vv{q})$ in \eref{modes:ansatz} so that 
they satisfy \eref{rte:rte}.
}
We normalize $\phi^m$ as
\bea
&&
\frac{1}{2\pi}\int_{\mathbb{S}^2}\phi^m(\nu,\mu(\hvv{k}))
\left(1-\mu(\hvv{k})^2\right)^{|m|}\rd{\hvv{s}}
\nonumber \\
&&=
\int_{-1}^1\phi^m(\nu,\mu)\rd{m(\mu)}
\nonumber \\
&&=
1.
\label{modes:normalization}
\eea
We will calculate singular eigenfunctions $\phi^m$ below.

By plugging (\ref{modes:ansatz}) into the radiative transport
equation (\ref{rte:rte}), we obtain
\bea
&&
\left(1-\frac{\mu(\hvv{k})}{\nu}\right)\phi^m(\nu,\mu(\hvv{k}))
\left(1-\mu(\hvv{k})^2\right)^{|m|/2}\re{\ri m\varphi(\hvv{k})}
\nonumber \\
&=&
c\int_{\mathbb{S}^2}f\left(\hvv{s}(\hvv{k})\cdot\hvv{s}'(\hvv{k})\right)
\phi^m(\nu,\mu'(\hvv{k}))\left(1-\mu'(\hvv{k})^2\right)^{|m|/2}
\nonumber \\
&\times&
\re{\ri m\varphi'(\hvv{k})}\rd{\hvv{s}'},
\label{rte:RTEansatz}
\eea
where we used $\hvv{s}\cdot\hvv{k}=\mu(\hvv{k})$ and expressed 
$f(\hvv{s}\cdot\hvv{s}')$ in the rotated reference frame.  The right-hand 
side is calculated as
\bea
\mbox{RHS}
&=&
2\pi c\Theta\left(N-|m|\right)\left(1-\mu(\hvv{k})^2\right)^{|m|/2}
\re{\ri m\varphi(\hvv{k})}
\nonumber \\
&\times&
\sum_{l'=|m|}^Nf_{l'}\frac{2l'+1}{4\pi}\frac{(l'-m)!}{(l'+m)!}
\nonumber \\
&\times&
p_{l'}^m(\mu(\hvv{k}))\int_{-1}^1p_{l'}^m(\mu')\phi^m(\nu,\mu')\rd{m(\mu')},
\eea
where the step function $\Theta(\cdot)$ is defined as 
$\Theta(x)=1$ for $x\ge0$ and $=0$ for $x<0$.  Hence,
\bea
\left(\nu-\mu(\hvv{k})\right)\phi^m(\nu,\mu(\hvv{k}))
=
2\pi c\nu\Theta\left(N-|m|\right)
\nonumber \\
\times
\sum_{l'=|m|}^Nf_{l'}\frac{2l'+1}{4\pi}
\frac{(l'-m)!}{(l'+m)!}p_{l'}^m(\mu(\hvv{k}))h_{l'}^m(\nu),
\label{phiisgiven}
\eea
where we defined
\be
h_l^m(\nu)=\int_{-1}^1\phi^m(\nu,\mu)p_l^m(\mu)\rd{m(\mu)}.
\ee
\mm{The polynomials $h_l^m$ were introduced by Mika \cite{Mika61} for $m=0$ 
and then generalized by McCormick and Ku\v{s}\v{c}er \cite{McCormick-Kuscer66} 
for general $m$.  
}
Since the right-hand side of \eref{phiisgiven} is zero for $|m|>N$ and then 
$\phi^m=0$, hereafter we suppose
\be
0\le|m|\le N.
\ee

Let us define
\be
\sigma_l=1-cf_l\Theta(N-l).
\ee
From \eref{phiisgiven}, we obtain
\bea
\sigma_l\nu h_l^m(\nu)=
\int_{-1}^1\mu\phi^m(\nu,\mu)p_l^m(\mu)\rd{m(\mu)}.
\label{rte:RTEansatzProj}
\eea
\Eref{rte:RTEansatzProj} implies the \mm{three-term} recurrence relation 
for $h_l^m(\nu)$ 
\cite{Inonu70}:
\be
\nu(2l+1)\sigma_lh_l^m(\nu)-(l-m+1)h_{l+1}^m(\nu)-(l+m)h_{l-1}^m(\nu)=0,
\label{recursion}
\ee
with
\be
h_{|m|}^m(\nu)=\mm{p_{|m|}^m},
\ee
and
\be
h_{|m|+1}^{|m|}(\nu)=(2|m|+1)\nu\sigma_{|m|}h_{|m|}^{|m|}(\nu).
\ee
We also have
\be
h_l^{-|m|}(\nu)=(-1)^{|m|}\frac{(l-|m|)!}{(l+|m|)!}h_l^{|m|}(\nu).
\ee
The functions $h_l^m(\nu)$ are computed using \eref{recursion}.

Let us define
\be
\mm{g}^m(\nu,\mu(\hvv{k}))
=\sum_{l'=|m|}^N(2l'+1)f_{l'}\frac{(l'-m)!}{(l'+m)!}p_{l'}^m(\mu(\hvv{k}))
h_{l'}^m(\nu).
\ee
We note that $\mm{g}^{-m}(\nu,\mu(\hvv{k}))=\mm{g}^m(\nu,\mu(\hvv{k}))$.  
The function $\phi^m$ is obtained as
\bea
\phi^m(\nu,\mu(\hvv{k}))
&=&
\frac{c\nu}{2}\mathcal{P}\frac{\mm{g}^m(\nu,\mu(\hvv{k}))}{\nu-\mu(\hvv{k})}
\nonumber \\
&+&
\lambda^m(\nu)(1-\nu^2)^{-|m|}\delta(\nu-\mu(\hvv{k})),
\label{eigenfunction}
\eea
where \mm{$\lambda^m(\nu)$ is given below.}

\section{Discrete eigenvalues and continuous spectrum}
\label{evals}

\mm{By multiplying $\left(1-\mu(\hvv{k})^2\right)^{|m|}$ and integrating 
over $\hvv{s}$, \eref{eigenfunction} becomes
}
\be
1=\frac{c\nu}{2}\pint_{-1}^1\frac{\mm{g}^m(\nu,\mu)}{\nu-\mu}\rd{m(\mu)}+
\int_{-1}^1\lambda^m(\nu)\delta(\nu-\mu)\rd{\mu}.
\ee

\mm{For $\nu\in(-1,1)$ we obtain}
\be
\lambda^m(\nu)
=
1-\frac{c\nu}{2}\pint_{-1}^1\frac{\mm{g}^m(\nu,\mu)}{\nu-\mu}\rd{m(\mu)}.
\ee
Note that $\lambda^{-m}(\nu)=\lambda^m(\nu)$ \mm{and hence} 
$\phi^{-m}(\nu,\mu(\hvv{k}))=\phi^m(\nu,\mu(\hvv{k}))$.

Let us define
\be
\Lambda^m(z)
=1-\frac{cz}{2}\int_{-1}^1\frac{\mm{g}^m(z,\mu)}{z-\mu}\rd{m(\mu)},
\ee
where $z\in\mathbb{C}$.  Eigenvalues $\nu\notin[-1,1]$ are solutions to 
\be
\Lambda^m(\nu)=0.
\label{rootsearch}
\ee
We write these discrete eigenvalues as $\pm\nu_j^m$ 
($\nu_0^m>\nu_1^m>\cdots>\nu_{M-1}^m>1$).  Note that $\nu_j^{-m}=\nu_j^m$.  
The number of discrete 
eigenvalues $M$ depends on $|m|$ and we have \cite{Mika61,McCormick-Kuscer66} 
$M\le N-|m|+1$.  
For $\nu\in(-1,1)$, we have the continuous spectrum.

\section{Method of rotated reference frames}
\label{mrrf}

Let us expand singular eigenfunctions with spherical harmonics.  
By introducing $c^m_l(\nu)$, we write
\be
\Phi^m_{\nu}(\hvv{s};\hvv{k})
=\sum_{l=|m|}^{\infty}c^m_l(\nu)Y_{lm}(\hvv{s};\hvv{k}).
\label{mrrf:eq1}
\ee
The calculation of the specific intensity by this expansion is called 
the method of rotated reference frames 
\cite{Markel04,Panasyuk06,Schotland-Markel07,Machida10}.  
From \eref{rte:RTEansatz}, we obtain
\bea
c^m_l(\nu)-\frac{1}{\nu}\sum_{l'=|m|}^{\infty}
\left(\int_{\mathbb{S}^2}\mu Y_{l'm}(\hvv{s})Y_{lm}^*(\hvv{s})\rd{\hvv{s}}
\right)c^m_{l'}(\nu)
\nonumber \\
=cf_l\Theta(N-l)c^m_l(\nu).
\eea
Hence we arrive at an eigenproblem:
\be
B^m|\psi^m(\nu)\rangle=\nu|\psi^m(\nu)\rangle,
\ee
where
\bea
B^m_{ll'}
&=&
\frac{1}{\sqrt{\sigma_l\sigma_{l'}}}
\int_{\mathbb{S}^2}\mu Y_{l'm}(\hvv{s})Y_{lm}^*(\hvv{s})\rd{\hvv{s}}
\nonumber \\
&=&
\sqrt{\frac{l^2-m^2}{(4l^2-1)\sigma_l\sigma_{l-1}}}\delta_{l',l-1}
\nonumber \\
&+&
\sqrt{\frac{(l+1)^2-m^2}{(4(l+1)^2-1)\sigma_{l+1}\sigma_l}}\delta_{l',l+1},
\eea
\be
\langle l|\psi^m(\nu)\rangle
=
\frac{1}{\sqrt{Z^m(\nu)}}\sqrt{\sigma_l}c^m_l(\nu),
\label{mrrf:eq2}
\ee
where the normalization \mm{factor} $Z^m(\nu)$ will be determined below so 
that $\langle\psi^m(\nu)|\psi^m(\nu)\rangle=1$ is satisfied.  
Note that \mm{$\Phi^m_{\nu}$ and $|\psi^m(\nu)\rangle$ are related as}
\be
\Phi^m_{\nu}(\hvv{s};\hvv{k})=
\sqrt{Z^m(\nu)}\sum_{l=|m|}^{\infty}
\frac{\langle l|\psi^m(\nu)\rangle}{\sqrt{\sigma_l}}Y_{lm}(\hvv{s};\hvv{k}).
\label{mrrf:evecs}
\ee

\mm{In Appendix \ref{efunc},} expansion coefficients $c^m_l(\nu)$ in 
\eref{mrrf:eq1} \mm{are} calculated 
using \eref{mrrf:eq2} and \eref{mrrf:evecs}.

\section{The Green's function}
\label{gfunc3d}

Let us consider the Green's function of the radiative transport equation 
in an infinite medium.  The Green's function obeys
\bea
&&
\hvv{s}\cdot\nabla G(\vv{r},\hvv{s};\vv{r}_0,\hvv{s}_0)+
G(\vv{r},\hvv{s};\vv{r}_0,\hvv{s}_0)
\nonumber \\
&=&
c\int_{\mathbb{S}^2}f(\hvv{s}\cdot\hvv{s}')
G(\vv{r},\hvv{s}';\vv{r}_0,\hvv{s}_0)\rd{\hvv{s}'}
\nonumber \\
&+&
\delta(\vv{r}-\vv{r}_0)\delta(\hvv{s}-\hvv{s}_0).
\label{defgreenfunc}
\eea
To proceed, we introduce $\tilde{\Phi}^m_{\nu}(\hvv{s};\hvv{k})$ such that
\be
\int_{\mathbb{S}^2}\mu\Phi^m_{\nu}(\hvv{s};\hvv{k})
\left[\tilde{\Phi}^m_{\nu'}(\hvv{s};\hvv{k}')\right]^*\rd{\hvv{s}}=
\delta_{\nu\nu'},
\ee
where the Kronecker delta $\delta_{\nu\nu'}$ is understood as the Dirac 
delta $\delta(\nu-\nu')$ for the continuous spectrum.  
The function $\tilde{\Phi}^m_{\nu}(\hvv{s};\hvv{k})$ will be determined 
as we compute the Green's function.  

We replace the source term in \eref{defgreenfunc} by a jump condition and solve
\begin{widetext}
\be
\bc
\hvv{s}\cdot\nabla G(\vv{r},\hvv{s};\vv{r}_0,\hvv{s}_0)+
G(\vv{r},\hvv{s};\vv{r}_0,\hvv{s}_0)=
c\int_{\mathbb{S}^2}f(\hvv{s}\cdot\hvv{s}')
G(\vv{r},\hvv{s}';\vv{r}_0,\hvv{s}_0)\rd{\hvv{s}'},
\\
G(\vv{\rho},z_0+0,\hvv{s};\vv{r}_0,\hvv{s}_0)-
G(\vv{\rho},z_0-0,\hvv{s};\vv{r}_0,\hvv{s}_0)=
\frac{1}{\hvv{s}\cdot\hvv{z}}
\delta(\vv{\rho}-\vv{\rho}_0)\delta\left(\hvv{s}-\hvv{s}_0\right),
\ec
\ee
\end{widetext}
with the boundary condition
$\lim_{|\vv{r}|\to\infty}G(\vv{r},\hvv{s};\vv{r}_0,\hvv{s}_0)=0$ and
$\vv{r}=(\vv{\rho},z)$ ($\vv{\rho}\in\mathbb{R}^2$, $z\in\mathbb{R}$), 
where $\vv{\rho}=(x,y)$.  Let us expand the Green's function using 
\mm{elementary solutions or} normal modes 
\eref{modes:ansatz}.
\begin{widetext}
\be
\bc
G(\vv{\rho},z,\hvv{s};\vv{\rho}_0,z_0,\hvv{s}_0)&=
\sum_{m=-N}^N
\int_{\mathbb{R}^2}\Biggl[
\sum_{j=0}^{M-1}a_{j+}^m(\vv{q})\mm{I}_{j+}^m(\vv{r},\hvv{s};\vv{q})
+
\int_0^1A_{\nu}^m(\vv{q})\mm{I}_{\nu}^m(\vv{r},\hvv{s};\vv{q})\rd{\nu}
\Biggr]\frac{{\rm d}\vv{q}}{(2\pi)^2},
\quad z>z_0,
\\
G(\vv{\rho},z,\hvv{s};\vv{\rho}_0,z_0,\hvv{s}_0)&=
-\sum_{m=-N}^N
\int_{\mathbb{R}^2}\Biggl[
\sum_{j=0}^{M-1}a_{j-}^m(\vv{q})\mm{I}_{j-}^m(\vv{r},\hvv{s};\vv{q})
+
\int_{-1}^0A_{\nu}^m(\vv{q})\mm{I}_{\nu}^m(\vv{r},\hvv{s};\vv{q})\rd{\nu}
\Biggr]\frac{{\rm d}\vv{q}}{(2\pi)^2},
\quad z<z_0.
\ec
\ee
\end{widetext}
From the jump condition, coefficients $a_{j\pm}^m$ and $A_{\nu}^m$ are 
determined as
\begin{widetext}
\be
a_{j\pm}^m(\vv{q})
=\re{-\ri\vv{q}\cdot\vv{\rho}_0}\re{\pm Q(\nu_j^m\mm{q})z_0/\nu_j^m}
\left[\tilde{\Phi}^m_{j\pm}(\hvv{s}_0;\hvv{k})\right]^*,
\quad
A_{\nu}^m(\vv{q})
=\re{-\ri\vv{q}\cdot\vv{\rho}_0}\re{Q(\nu \mm{q})z_0/\nu}
\left[\tilde{\Phi}^m_{\nu}(\hvv{s}_0;\hvv{k})\right]^*.
\ee
\end{widetext}
Hence the Green's function is written as
\begin{widetext}
\bea
G(\vv{\rho},z,\hvv{s};\vv{\rho}_0,z_0,\hvv{s}_0)
&=&
\frac{\pm 1}{(2\pi)^2}\int_{\mathbb{R}^2}
\re{\ri\vv{q}\cdot(\vv{\rho}-\vv{\rho}_0)}\sum_{m=-N}^N
\Biggl\{
\sum_{j=0}^{M-1}\Phi^m_{j\pm}(\hvv{s};\hvv{k})
\left[\tilde{\Phi}^m_{j\pm}(\hvv{s}_0;\hvv{k})\right]^*
\re{-Q(\nu_j^m\mm{q})|z-z_0|/\nu_j^m}
\nonumber \\
&+&
\int_0^1\Phi^m_{\pm\nu}(\hvv{s};\hvv{k})
\left[\tilde{\Phi}^m_{\pm\nu}(\hvv{s}_0;\hvv{k})\right]^*
\re{-Q(\nu\mm{q})|z-z_0|/\nu}\rd{\nu}
\Biggr\}
\rd{\vv{q}},
\label{green2}
\eea
\end{widetext}
where upper signs are chosen for $z>z_0$ and lower signs are chosen for 
$z<z_0$.

\mm{To find $\tilde{\Phi}^m_{\nu}$, we note that} the Green's function 
obtained with the method of rotated reference frames \cite{Panasyuk06} is 
expressed as
\bea
&&
G(\vv{r},\hvv{s};\vv{r}_0,\hvv{s}_0)
=
\frac{1}{(2\pi)^2}\int_{\mathbb{R}^2}\re{\ri\vv{q}\cdot(\vv{\rho}-\vv{\rho}_0)}
\sum_{\nu>0}\sum_{m=-N}^N
\nonumber \\
&&
\frac{1}{\nu Q(\nu\mm{q})Z^m(\nu)}\Phi^m_{\pm\nu}(\hvv{s};\hvv{k})
\left[\Phi^m_{\pm\nu}(\hvv{s}_0;\hvv{k})\right]^*
\re{-Q(\nu\mm{q})|z-z_0|/\nu}\rd{\vv{q}}\mm{,}
\nonumber \\
\label{george}
\eea
\mm{where we used the relation \eref{mrrf:evecs}}.  
By comparing \eref{green2} and \eref{george}, we obtain
\be
\tilde{\Phi}^m_{\pm\nu}(\hvv{s};\hvv{k})
=\Phi^m_{\pm\nu}(\hvv{s};\hvv{k})
\left[\pm\nu Q(\nu \mm{q})Z^m(\nu)\right]^{-1}.
\ee

To determine $Z^m(\nu)$, we consider the one-dimensional case.  
By integrating the Green's function over $\vv{\rho}_0$, we obtain 
\bea
G(z,\hvv{s};z_0,\hvv{s}_0)
&=&
\pm\sum_{m=-N}^N\Biggl\{
\nonumber \\
&&
\sum_{j=0}^{M-1}\Phi^m_{j\pm}(\hvv{s};\hvv{z})
\left[\tilde{\Phi}^m_{j\pm}(\hvv{s}_0;\hvv{z})\right]^*
\re{-|z-z_0|/\nu_j^m}
\nonumber \\
&+&
\int_0^1\Phi^m_{\pm\nu}(\hvv{s};\hvv{z})
\left[\tilde{\Phi}^m_{\pm\nu}(\hvv{s}_0;\hvv{z})\right]^*
\re{-|z-z_0|/\nu}\rd{\nu}
\Biggr\}.
\nonumber \\
\label{green3}
\eea
On the other hand, the one-dimensional Green's function is given by 
\cite{Mika61,McCormick-Kuscer66}
\begin{widetext}
\bea
G(z,\hvv{s};z_0,\hvv{s}_0)
&=&
\frac{1}{2\pi}\sum_{m=-N}^N\Biggl\{
\sum_{j=0}^{M-1}\frac{1}{\mathcal{N}^m_j}
\phi^m(\pm\nu_j^m,\mu)\phi^m(\pm\nu_j^m,\mu_0)
\left(1-\mu^2\right)^{|m|}
\mm{\re{\ri m(\varphi-\varphi_0)}}
\re{-|z-z_0|/\nu_j^m}
\nonumber \\
&+&
\int_0^1\frac{1}{\mathcal{N}^m(\nu)}\phi^m(\pm\nu,\mu)\phi^m(\pm\nu,\mu_0)
\left(1-\mu^2\right)^{|m|}
\mm{\re{\ri m(\varphi-\varphi_0)}}
\re{-|z-z_0|/\nu}\rd{\nu}
\Biggr\},
\label{green1d}
\eea
\end{widetext}
where
\be
\mathcal{N}^m_j
\mm{=\mathcal{N}^m(\nu_j^m)}
=\frac{\mm{c}}{2}(\nu_j^m)^2\mm{g}(\nu_j^m,\nu_j^m)
\frac{{\rm d}\Lambda^m(z)}{{\rm d}z}\Biggm|_{z=\nu_j^m},
\ee
\mm{and for $\nu\in(-1,1)$,}
\be
\mathcal{N}^m(\nu)
=\nu\Lambda^{m+}(\nu)\Lambda^{m-}(\nu)\left(1-\nu^2\right)^{-|m|}.
\ee
Here $\Lambda^{m\pm}(\nu)=\lim_{\epsilon\to0+}\Lambda^m(\nu\pm\ri\epsilon)$.  
By comparing \eref{green3} and \eref{green1d}, we obtain
\be
\nu_j^mZ^m(\nu_j^m)=2\pi\mathcal{N}^m_j,\quad
\nu Z^m(\nu)=2\pi\mm{\mathcal{N}^m}(\nu),
\ee
where $\nu$ belongs to the continuous spectrum.  
Finally we obtain
\mm{
\be
\tilde{\Phi}^m_{\nu}(\hvv{s};\hvv{k})
=\Phi^m_{\nu}(\hvv{s};\hvv{k})
\left[2\pi Q(\nu q)\mathcal{N}^m(\nu)\right]^{-1},
\ee
where $\nu=\pm\nu_j^m$ or $\nu\in(-1,1)$.
}
The Green's function \mm{is obtained} as
\begin{widetext}
\bea
G(\vv{\rho},z,\hvv{s};\vv{\rho}_0,z_0,\hvv{s}_0)
&=&
\frac{1}{(2\pi)^3}\int_{\mathbb{R}^2}
\re{\ri\vv{q}\cdot(\vv{\rho}-\vv{\rho}_0)}\sum_{m=-N}^N
\Biggl\{
\sum_{j=0}^{M-1}\frac{1}{Q(\nu_j^m\mm{q})\mathcal{N}_j^m}
\Phi^m_{j\pm}(\hvv{s};\hvv{k})
\left[\Phi^m_{j\pm}(\hvv{s}_0;\hvv{k})\right]^*
\re{-Q(\nu_j^m\mm{q})|z-z_0|/\nu_j^m}
\nonumber \\
&+&
\int_0^1\frac{1}{Q(\nu\mm{q})\mm{\mathcal{N}^m}(\nu)}
\Phi^m_{\pm\nu}(\hvv{s};\hvv{k})
\left[\Phi^m_{\pm\nu}(\hvv{s}_0;\hvv{k})\right]^*
\re{-Q(\nu\mm{q})|z-z_0|/\nu}\rd{\nu}
\Biggr\}
\rd{\vv{q}}.
\label{finalgreenfunc}
\eea
\end{widetext}

\mm{As the simplest case, let us consider} the isotropic scattering $N=0$
\mm{. We then have}
\be
\Phi^0_{\nu}(\hvv{s};\hvv{k})
=
\frac{c\nu}{2}\mathcal{P}\frac{1}{\nu-\mu(\hvv{k})}+
\lambda^0(\nu)\delta(\nu-\mu(\hvv{k}))
=\phi^{\mm{0}}(\nu,\mu(\hvv{k})),
\ee
where 
\be
\lambda^0(\nu)=1-\frac{c\nu}{2}\pint_{-1}^1\frac{1}{\nu-\mu}\rd{\mu}
=1-c\nu\tanh^{-1}{\nu}.
\ee
In this case $M=1$ and the discrete eigenvalues $\pm\nu^0_0=\pm\nu_0$ 
are solutions to
\be
\Lambda^0(z)=1-\frac{cz}{2}\int_{-1}^1\frac{1}{z-\mu}\rd{\mu}
=1-cz\tanh^{-1}\frac{1}{z}=0.
\ee
The Green's function is obtained as
\begin{widetext}
\bea
G(\vv{\rho},z,\hvv{s};\vv{\rho}_0,z_0,\hvv{s}_0)
&=&
\frac{1}{(2\pi)^3}\int_{\mathbb{R}^2}
\re{\ri\vv{q}\cdot(\vv{\rho}-\vv{\rho}_0)}
\Biggl\{
\frac{1}{Q(\nu_0\mm{q})\mathcal{N}_0}
\phi^{\mm{0}}\left(\pm\nu_0,\mu(\hvv{k})\right)
\phi^{\mm{0}*}\left(\pm\nu_0,\mu_0(\hvv{k})\right)
\re{-Q(\nu_0\mm{q})|z-z_0|/\nu_0}
\nonumber \\
&+&
\int_0^1\frac{1}{Q(\nu\mm{q})\mathcal{N}(\nu)}
\phi^{\mm{0}}\left(\pm\nu,\mu(\hvv{k})\right)
\phi^{\mm{0}*}\left(\pm\nu,\mu_0(\hvv{k})\right)
\re{-Q(\nu\mm{q})|z-z_0|/\nu}\rd{\nu}
\Biggr\}
\rd{\vv{q}}.
\nonumber \\
\label{gfunciso}
\eea
\end{widetext}
\mm{If we integrate \eref{gfunciso} with respect to $\vv{\rho}_0$, 
$G$ in \eref{gfunciso} becomes the one-dimensional Green's function 
written in the book by Case and Zweifel \cite{Case-Zweifel}.
}

\section{\mm{Energy density}}
\label{den}

\mm{Let us calculate the energy density $U$.  For simplicity, we assume 
linear scattering, $N=1$.  we measure $U$ along the $z$-axis.  
The energy density $U$ is given by
\be
U(z)=\frac{1}{v}\int_{\mathbb{S}^2}I(\vv{\rho}=\vv{0},z,\hvv{s})\rd{\hvv{s}},
\ee
where $v$ is the speed of light in the medium and $I$ is 
the specific intensity obeying \eref{rte:fullrte1}.
}

\begin{figure}[htbp]
\centerline{\includegraphics[width=0.9\columnwidth]{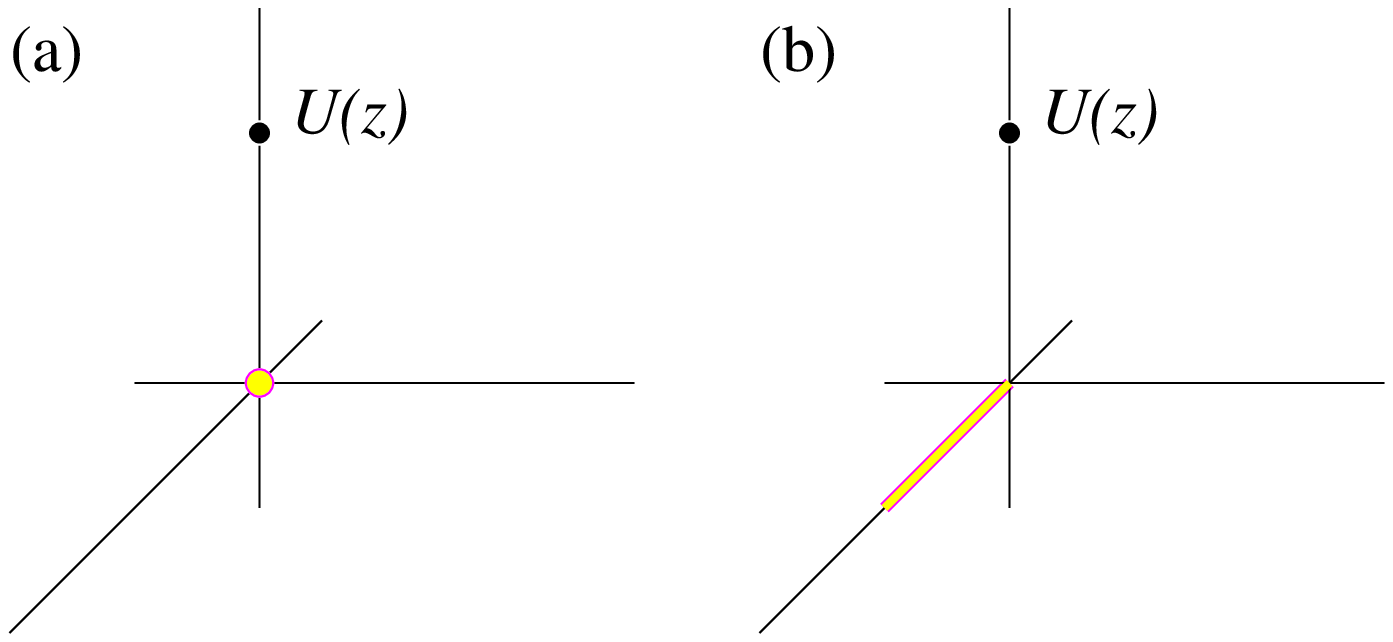}}
\caption{
\mm{(Color online) The energy density $U(z)$ for (a) the point source and 
(b) the source of length $\ell$ on the $x$-axis.}
}
\label{pic}
\end{figure}

\begin{figure}[htbp]
\centerline{\includegraphics[width=1.0\columnwidth]{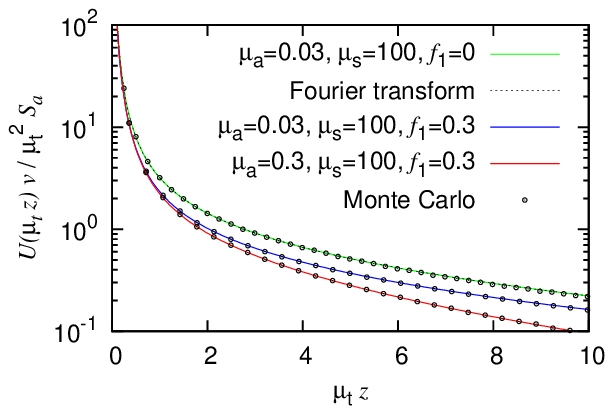}}
\caption{
\mm{(Color online) The energy density \eref{Unu} is plotted together with 
\eref{den:Ufourier1} and results from Monte Carlo simulation.  
The optical parameters $(\mu_{\rm a},\,\mu_{\rm s},\,f_1)$ are, 
from the top, $(0.03\,{\rm cm}^{-1},\,100\,{\rm cm}^{-1},\,0)$, 
$(0.03\,{\rm cm}^{-1},\,100\,{\rm cm}^{-1},\,0.3)$, and 
$(0.3\,{\rm cm}^{-1},\,100\,{\rm cm}^{-1},\,0.3)$.
} 
}
\label{fig1}
\end{figure}

\mm{First we place an isotropic source at the origin, 
$S=S_a\delta(\vv{r})$ with constant $S_a$ in \eref{rte:fullrte1} 
(see \fref{pic}(a)).  We have
\bea
I(\vv{0},z,\hvv{s})
&=&
\mu_{\rm t}^2S_a\int_{\mathbb{R}^3\times\mathbb{S}^2}
G(\vv{0},z,\hvv{s};\vv{\rho}_0,z_0,\hvv{s}_0)
\nonumber \\
&\times&
\delta(\vv{\rho}_0)\delta(z_0)d\vv{\vv{\rho}}_0dz_0d\hvv{s}_0,
\eea
where $z,\vv{\rho},z_0$ are measured in the unit of $1/\mu_{\rm t}$.  
In this case, $U$ is spherically symmetric.  Using \eref{finalgreenfunc} 
we obtain
\bea
\frac{U(z)}{\mu_{\rm t}^2S_a}
=
\int_{\mathbb{S}^2\times\mathbb{S}^2}
G(\vv{0},z,\hvv{s};\vv{0},0,\hvv{s}_0)\rd{\hvv{s}}\rd{\hvv{s}_0}
\nonumber \\
=
\frac{1}{vz}\left[
\frac{\re{-z/\nu_0}}{\nu_0\mathcal{N}_0}
+\int_0^1\frac{\re{-z/\nu}}{\nu\mathcal{N}(\nu)}\rd{\nu}\right],
\quad z>0.
\label{Unu}
\eea
Here $\nu_0$ is the positive solution to $\Lambda^0(\nu_0)=0$, where
\be
\Lambda^0(\nu_0)=
1-\frac{c\nu_0}{2}\int_{-1}^1\frac{g^0(\nu_0,\mu)}{\nu_0-\mu}\rd{\mu}.
\label{den:Lambda}
\ee
We consider the following three cases: 
(i) $\mu_{\rm a}=0.03\,{\rm cm}^{-1}$, $\mu_{\rm s}=100\,{\rm cm}^{-1}$, 
$\mm{f_1}=0$ ($c=0.9997$), 
(ii) $\mu_{\rm a}=0.03\,{\rm cm}^{-1}$, $\mu_{\rm s}=100\,{\rm cm}^{-1}$, 
$\mm{f_1}=0.3$ ($c=0.9997$) \cite{Michels08}, 
and (iii) 
$\mu_{\rm a}=0.3\,{\rm cm}^{-1}$, $\mu_{\rm s}=100\,{\rm cm}^{-1}$, 
$\mm{f_1}=0.3$ ($c=0.997$).  
In the case (i) with $f_1=0$, the density can also be obtained with 
the Fourier transform.  The Green's function is obtained as 
\bea
G(\vv{r},\hvv{s};\vv{r}_0,\hvv{s}_0)
=
G_0(\vv{r},\hvv{s};\vv{r}_0,\hvv{s}_0)
+\frac{c}{4\pi(2\pi)^3}\int_{\mathbb{R}^3}
\re{\ri\vv{k}\cdot(\vv{r}-\vv{r}_0)}
\nonumber \\
\times
\frac{1}{(1+\ri\vv{k}\cdot\hvv{s})(1+\ri\vv{k}\cdot\hvv{s}_0)}
\left[1-\frac{c}{|\vv{k}|}\tan^{-1}\left(|\vv{k}|\right)\right]^{-1}
\rd{\vv{k}},
\nonumber \\
\label{den:gfuncfourieir}
\eea
where
\be
G_0(\vv{r},\hvv{s};\vv{r}_0,\hvv{s}_0)
=
\frac{\re{-|\vv{r}-\vv{r}_0|}}{|\vv{r}-\vv{r}_0|^2}
\delta\left(\hvv{s}-\frac{\vv{r}-\vv{r}_0}{|\vv{r}-\vv{r}_0|}\right)
\delta(\hvv{s}-\hvv{s}_0).
\label{den:gfuncfourieir0}
\ee
Using \eref{den:gfuncfourieir} and \eref{den:gfuncfourieir0}, we obtain
\be
\frac{U(z)}{\mu_{\rm t}^2S_a}
=\frac{\re{-|z|}}{vz^2}
+\frac{2c}{\pi v}\int_0^{\infty}\frac{\sin(kz)}{z}
\frac{\left(\tan^{-1}k\right)^2}{k-c\tan^{-1}k}\rd{k}.
\label{den:Ufourier1}
\ee
In \fref{fig1}, we plot $U(z)v/\mu_{\rm t}^2S_a$ as a function of $z$.  
In addition to \eref{Unu}, densities by \eref{den:Ufourier1} and by Monte Carlo 
simulation are shown.  We see perfect agreement.
}

\begin{figure}[htbp]
\centerline{\includegraphics[width=1.0\columnwidth]{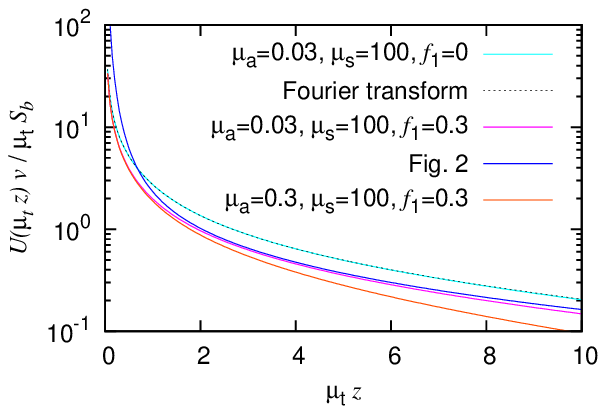}}
\caption{
\mm{(Color online) The energy density \eref{den:Unuax} is plotted.  
The optical parameters 
$(\mu_{\rm a},\,\mu_{\rm s},\,f_1)$ are, 
from the top, $(0.03\,{\rm cm}^{-1},\,100\,{\rm cm}^{-1},\,0)$, 
$(0.03\,{\rm cm}^{-1},\,100\,{\rm cm}^{-1},\,0.3)$, and 
$(0.3\,{\rm cm}^{-1},\,100\,{\rm cm}^{-1},\,0.3)$.  
The blue line from \fref{fig1} shows \eref{Unu} for 
$(\mu_{\rm a},\,\mu_{\rm s},\,f_1)=
(0.03\,{\rm cm}^{-1},\,100\,{\rm cm}^{-1},\,0.3)$.  
}
}
\label{fig2}
\end{figure}

\mm{Next we consider the source of length $\ell$ on the $x$-axis 
(see \fref{pic}(b)), i.e., we put 
$S=S_b\Theta(\ell-x)\Theta(x)\delta(y)\delta(z)$ with constant $S_b$ 
in \eref{rte:fullrte1}.  We have
\bea
&&
I(\vv{0},z,\hvv{s})
=
\mu_{\rm t}S_b\int_{\mathbb{R}^3\times\mathbb{S}^2}
G(\vv{0},z,\hvv{s};x_0,y_0,z_0,\hvv{s}_0)
\nonumber \\
&&
\times
\Theta(\mu_{\rm t}\ell-x_0)\Theta(x_0)\delta(y_0)\delta(z_0)
dx_0dy_0dz_0d\hvv{s}_0,
\eea
where $z,x_0,y_0,z_0$ are measured in the unit of $1/\mu_{\rm t}$.  
We compute $I$ using \eref{finalgreenfunc} and obtain
\bea
\frac{U(z)}{\mu_{\rm t}S_b}
&=&
\int_{\mathbb{R}^2}\int_{\mathbb{S}^2\times\mathbb{S}^2}
G(\vv{0},z,\hvv{s};\vv{\rho}_0,0,\hvv{s}_0)
\nonumber \\
&\times&
\Theta(\mu_{\rm t}\ell-x_0)\Theta(x_0)\delta(y_0)
\rd{\hvv{s}}\rd{\hvv{s}_0}\rd{\vv{\rho}_0}
\nonumber \\
&=&
\frac{1}{v}\int_0^{\infty}\left(\int_0^{\mu_{\rm t}\ell q}J_0(t)\rd{t}\right)
\Biggl[\frac{\re{-Q(\nu_0q)z/\nu_0}}{Q(\nu_0q)\mathcal{N}_0}
\nonumber \\
&+&
\int_0^1\frac{\re{-Q(\nu q)z/\nu}}{Q(\nu q)\mathcal{N}(\nu)}\rd{\nu}
\Biggr]\rd{q},
\quad z>0,
\label{den:Unuax}
\eea
where $\nu_0$ is the positive root of \eref{den:Lambda} and $J_0(u)$ 
is the zeroth order Bessel function of the first kind.  In addition, 
with the Fourier transform, we obtain
\bea
\frac{U(z)}{\mu_{\rm t}S_b}
&=&
\frac{1}{v}\int_0^{\mu_{\rm t}\ell}\Biggl[
\frac{\re{-\sqrt{x_0^2+z^2}}}{x_0^2+z^2}
\nonumber \\
&+&
\frac{2c}{\pi}\int_0^{\infty}
\frac{\sin(k\sqrt{x_0^2+z^2})}{\sqrt{x_0^2+z^2}}
\frac{\left(\tan^{-1}k\right)^2}{k-c\tan^{-1}k}\rd{k}
\Biggr]
\rd{x_0}.
\nonumber \\
\label{den:Ufourier2}
\eea
Let us put $\mu_{\rm t}\ell=1$.  
In \fref{fig2}, we plot \eref{den:Unuax} together with \eref{den:Ufourier2}.  
Moreover \eref{Unu} for $(\mu_{\rm a},\,\mu_{\rm s},\,f_1)=
(0.03\,{\rm cm}^{-1},\,100\,{\rm cm}^{-1},\,0.3)$ is plotted for comparison.  
We see that $U$ is similar to the density in \fref{fig1} except for small $z$.
}

\section{Summary}
\label{summary}

We have constructed elementary solutions of the radiative transport 
equation 
\mm{in three dimensions.  Each elementary solution} 
carries the wave vector $\vv{k}$, and 
is labeled by Case's discrete eigenvalues and continuous spectrum.  
\mm{By virtue of rotated reference frames, the angular part of each 
elementary solution is given by the singular eigenfunction for the 
one-dimensional radiative transport equation.}

\mm{Using the elementary solutions, the Green's function in an infinite medium 
is obtained.  Moreover the energy density is computed for 
different sources and optical parameters.
}

\mm{
\section*{Acknowledgments}
Monte Carlo simulations were carried out using the package MC written by 
Vadim A. Markel 
(http://whale.seas.upenn.edu/vmarkel/CODES/MC.html).  
The original code was partially modified for linear scattering.
}

\appendix

\section{Polar and azimuthal angles in rotated reference frames}
\label{rotmuphi}

Let $\theta$ and $\varphi$ be the polar and azimuthal angles of $\hvv{s}$ 
in the laboratory frame.  
Let $\varphi_{\hvv{k}}$ and $\theta_{\hvv{k}}$ be the polar and azimuthal 
angles of $\hvv{k}$ in the laboratory frame.  
For $\hvv{k}=\left(-\ri\nu\vv{q},\,Q(\nu \mm{q})\right)$, we obtain
\be
\cos\theta_{\hvv{k}}=\hvv{k}\cdot\hvv{z}=Q(\nu \mm{q}),\quad
\sin\theta_{\hvv{k}}=\sqrt{1-\cos^2\theta_{\hvv{k}}}=\ri|\nu\vv{q}|,
\ee
and
\be
\varphi_{\hvv{k}}=
\bc
\varphi_{\vv{q}}+\pi
&\quad\mbox{for}\;\nu>0,
\\
\varphi_{\vv{q}}
&\quad\mbox{for}\;\nu<0,
\ec
\ee
where $\varphi_{\vv{q}}$ is the angle of $\vv{q}$.  Therefore, we have
\be
\mu(\hvv{k})
=\hvv{s}\cdot\hvv{k}
=-\ri\nu \mm{q}\sin\theta\cos(\varphi-\varphi_{\vv{q}})+
Q(\nu \mm{q})\cos\theta.
\label{rotatedmu}
\ee

In general, we can rotate functions as follows.  Let us introduce
rotated spherical harmonics $Y_{lm}(\hvv{s};\hvv{k})$ \cite{Markel04}:
\be
Y_{lm}(\hvv{s};\hvv{k})=
\mathcal{D}(\hvv{k})Y_{lm}(\hvv{s})=
\sum_{m'=-l}^lD_{m'm}^l(\varphi_{\hvv{k}},\theta_{\hvv{k}},0)Y_{lm'}(\hvv{s}),
\ee
where $D_{m'm}^l(\varphi_{\hvv{k}},\theta_{\hvv{k}},0)=
\re{-\ri m'\varphi_{\hvv{k}}}d_{m'm}^l(\theta_{\hvv{k}})$.  Here 
$d_{m'm}^l$ are the Wigner $d$-matrices \cite{VMK}.  That is, 
$Y_{lm}(\hvv{s},\hvv{k})$ are spherical harmonics defined in a 
rotated reference frame whose $z$-axis coincides with the direction 
of the unit vector $\hvv{k}$.  We have 
$Y_{lm}(\hvv{s})=Y_{lm}(\hvv{s};\hvv{z})$.  We write analytically 
continued Wigner's $d$-matrices as
\be
d_{m'm}^l(\theta_{\hvv{k}})=d_{m'm}^l[\ri\tau(\nu \mm{q})].
\ee
First a few matrices are obtained as
\be
d^0_{00}=1,
\ee
\be
d^1_{00}=\sqrt{1+x^2},\quad
d^1_{01}=\frac{\ri}{\sqrt{2}}|x|,\quad
d^1_{1\pm 1}=\frac{1\pm\sqrt{1+x^2}}{2}.
\ee
We note that $d^l_{mm'}=(-1)^{m+m'}d^l_{-m-m'}=(-1)^{m+m'}d^l_{m'm}$.  
All $d_{m'm}^l[\ri\tau(\nu \mm{q})]$ are computed using the recurrence 
relations \cite{Machida10}.  We obtain
\bea
\re{\ri m\varphi(\hvv{k})}
&=&
\left(1-\mu(\hvv{k})^2\right)^{-\mm{|}m\mm{|}/2}
\frac{(-1)^m\sqrt{4\pi(2m+1)!}}{(2m+1)!!}
\nonumber \\
&\times&
\sum_{m'=-m}^m\re{-\ri m'\varphi_{\hvv{k}}}d^m_{m'm}(\theta_{\hvv{k}})
Y_{mm'}(\hvv{s}),
\label{a:rotatedvarphi}
\eea
where $\theta$ satisfies $\cos\theta=\mu$ with $\mu$ in \eref{modes:ansatz}.

\section{Expansion coefficients}
\label{efunc}

Here we calculate $c^m_l(\nu)$.  We have
\bea
c^m_l(\nu)
&=&
\int_{\mathbb{S}^2}\left[
\frac{c\nu}{2}\mathcal{P}\frac{\mm{g}^m(\nu,\mu)}{\nu-\mu}+
\lambda^m(\nu)\left(1-\nu^2\right)^{-|m|}\delta(\nu-\mu)\right]
\nonumber \\
&\times&
\left(1-\mu^2\right)^{|m|/2}
\re{\ri m\varphi}Y_{lm}^*(\hvv{s})\rd{\hvv{s}}.
\eea
Hence,
\bea
c^m_l(\nu)
&=&
2\pi\sqrt{\frac{2l+1}{4\pi}\frac{(l-m)!}{(l+m)!}}\Biggl[
\frac{c\nu}{2}\sum_{l''=|m|}^N
f_{l''}(2l''+1)
\nonumber \\
&\times&
\frac{(l''-m)!}{(l''+m)!}h_{l''}^m(\nu)
(-1)^m\pint_{-1}^1\frac{P_{l''}^m(\mu)P_l^m(\mu)}{\nu-\mu}\rd{\mu}
\nonumber \\
&+&
\lambda^m(\nu)\left(1-\nu^2\right)^{-|m|/2}P_l^m(\nu)
\int_{-1}^1\delta(\nu-\mu)\rd{\mu}
\Biggr].
\nonumber \\
\eea
Note that $c^m_l(-\nu)=(-1)^{l+m}c^m_l(\nu)$ because 
$P_l^m(-\nu)=(-1)^{l+m}P_l^m(\nu)$.  
Therefore, we obtain for $\nu\notin[-1,1]$
\bea
c^m_l(\nu)
&=&
2\pi\sqrt{\frac{2l+1}{4\pi}\frac{(l-m)!}{(l+m)!}}
\frac{c\nu}{2}\sum_{l''=|m|}^Nf_{l''}(2l''+1)
\nonumber \\
&\times&
\frac{(l''-m)!}{(l''+m)!}h_{l''}^m(\nu)
(-1)^m2Q_{\max(l,l'')}^m(\nu)P_{\min(l,l'')}^m(\nu),
\nonumber \\
\eea
where $Q_{\max(l,l'')}^m(\nu)$ and $P_{\min(l,l'')}^m(\nu)$ have 
a branch cut from $-\infty$ to $1$ \cite{Hurwitz-Zweifel55}, 
and for $\nu\in(-1,1)$
\bea
c^m_l(\nu)
&=&
2\pi\sqrt{\frac{2l+1}{4\pi}\frac{(l-m)!}{(l+m)!}}\Biggl[
\frac{c\nu(-1)^m}{2}\sum_{l''=|m|}^N
f_{l''}(2l''+1)
\nonumber \\
&\times&
\frac{(l''-m)!}{(l''+m)!}h_{l''}^m(\nu)
\nonumber \\
&\times&
\left(
-\ri\pi P_{l''}^m(\nu)P_l^m(\nu)-\int_0^{\pi}
\frac{P_{l''}^m(\re{\ri\theta})P_l^m(\re{\ri\theta})}{\nu-\re{\ri\theta}}
\ri\re{\ri\theta}\rd{\theta}\right)
\nonumber \\
&+&
\lambda^m(\nu)(1-\nu^2)^{-|m|/2}P_l^m(\nu)\Biggr].
\eea

\end{document}